%% file: main.tex
\documentclass[sigconf]{acmart}

\usepackage{booktabs}
\usepackage{multirow}
\usepackage{graphicx}
\usepackage{xcolor}
\usepackage{amsmath}
\usepackage{balance}
\usepackage{enumitem}
\usepackage{tikz}
\usetikzlibrary{positioning,shapes.geometric,arrows.meta,fit}

\copyrightyear{2026}
\acmYear{2026}
\setcopyright{cc}
\setcctype{by}
\acmConference[SIGIR '26]{Proceedings of the 49th International ACM SIGIR Conference on Research and Development in Information Retrieval}{July 20--24, 2026}{Melbourne, VIC, Australia}
\acmBooktitle{Proceedings of the 49th International ACM SIGIR Conference on Research and Development in Information Retrieval (SIGIR '26), July 20--24, 2026, Melbourne, VIC, Australia}
\acmDOI{10.1145/3805712.3808561}
\acmISBN{979-8-4007-2599-9/2026/07}

\begin{document}

\title{Reproduction Beyond Benchmarks: ConstBERT and ColBERT-v2 Across Backends and Query Distributions}

\author{Utshab Kumar Ghosh}
\affiliation{%
  \institution{Missouri University of Science and Technology}
  \city{Rolla}
  \state{MO}
  \country{USA}
}
\orcid{0000-0003-3096-6909}
\email{u.ghosh@mst.edu}

\author{Ashish David}
\affiliation{%
  \institution{Missouri University of Science and Technology}
  \city{Rolla}
  \state{MO}
  \country{USA}
}
\orcid{0000-0003-3316-838X}
\email{ashishdavid@mst.edu}

\author{Shubham Chatterjee}
\affiliation{%
  \institution{Missouri University of Science and Technology}
  \city{Rolla}
  \state{MO}
  \country{USA}
}
\orcid{0000-0002-6729-1346}
\email{shubham.chatterjee@mst.edu}

\renewcommand{\shortauthors}{Ghosh, David, and Chatterjee}

\input{tex_files/abstract}



\maketitle

\input{tex_files/sections/intro}

\input{tex_files/sections/related-work}

\input{tex_files/sections/overview-of-original-work}

\input{tex_files/sections/experimental-setup}

\input{tex_files/sections/results}

\input{tex_files/sections/conclusion}

\bibliographystyle{ACM-Reference-Format}
\balance
\bibliography{references}

\end{document}

%% file: tex_files/abstract.tex
\begin{abstract}
Reproducibility must validate architectural robustness, not just numerical accuracy. We evaluate ColBERT-v2 and ConstBERT across five dimensions, finding that while ConstBERT reproduces within 0.05\% MRR@10 on MS-MARCO, both models show a drop of 86–97\% on long, narrative queries (TREC ToT 2025). Ablations prove this failure is architectural: performance plateaus at 20 words because the MaxSim operator's uniform token weighting cannot distinguish signal from filler noise. Furthermore, undocumented backend parameters create an 8-point gap due to ConstBERT's sparse centroid coverage, and fine-tuning with 3$\times$ more data actually degrades performance by up to 29\%. We conclude that architectural constraints in multi-vector retrieval cannot be overcome by adaptation alone. Code: \url{https://github.com/utshabkg/multi-vector-reproducibility}.
\end{abstract}

%% file: tex_files/sections/intro.tex
\section{Introduction}
\label{sec:Introduction}

Multi-vector retrieval models achieve state-of-the-art effectiveness on standard benchmarks, with ColBERT~\cite{khattab2020colbert} and ConstBERT~\cite{macavaney2025constbert} reaching approximately 39\% MRR@10 on MS-MARCO passage ranking. Unlike traditional dense retrieval that represents each text as a single vector, multi-vector models encode queries and documents as sets of embeddings---typically one per token---enabling fine-grained term-level matching through late interaction. This architectural design has demonstrated strong generalization across diverse domains: evaluations on BEIR~\cite{thakur2021beir}---spanning biomedical, financial, and scientific tasks---show these models maintain 47--50\% nDCG@10 across 13 datasets.

However, BEIR's datasets share a critical structural characteristic with MS-MARCO: short, well-formed queries (median 5--15 words). This raises an important question: \emph{Do multi-vector models' architectural advantages generalize to query distributions with fundamentally different structural properties?}\footnote{Throughout this paper, we follow the terminology used in the SIGIR Reproducibility Track call, where \emph{reproducibility} refers to evaluation by a different team under different experimental setups (termed \emph{replicability} in the revised ACM definitions).} This question extends beyond traditional reproducibility concerns about matching reported numbers---it asks whether the \emph{architectural benefits} claimed in original work (fine-grained matching, richer semantic interactions) hold when query characteristics differ substantially from training data. We reframe reproducibility as a diagnostic mechanism for understanding \emph{where and why} architectures succeed or fail, rather than merely confirming \emph{whether} reported numbers can be matched.

We evaluate two architecturally distinct multi-vector models: \textbf{ColBERT-v2}~\cite{santhanam2022colbertv2}, which utilizes variable-length representations, and \textbf{ConstBERT}~\cite{macavaney2025constbert}, which pools embeddings into exactly 32 vectors per document. Our study comprises: (1) \textit{Systematic Reproduction} across in-domain benchmarks (MS-MARCO), retrieval backends (FAISS-IVF vs.\ PLAID), and out-of-domain evaluation (BEIR); and (2) \textit{Generalization Stress Testing} on TREC Tip-of-the-Tongue (ToT)~\cite{trec2025tot}, featuring queries with extreme structural shifts (121-word narratives vs.\ 6-word factoids).

ToT represents an extreme distributional shift: queries are long narrative descriptions of half-remembered entities (median 121 words vs.\ MS-MARCO's 6), featuring ambiguous language (``I think it was about...''), hedging (``maybe''), and fragmented memories rather than well-formed questions. Such verbose queries arise naturally in practice. For example, voice-based interfaces promote conversational, natural-language input~\cite{SA202140}. ToT thus represents a stress test of architectural assumptions under realistic verbose query scenarios that standard benchmarks systematically exclude. These queries stress-test the MaxSim operator---the core scoring function in multi-vector retrieval that sums maximum similarities across all query tokens. ToT postdates both models and was not used in original evaluations, enabling us to test whether claimed architectural advantages generalize beyond benchmark-specific query distributions.

\smallskip

\noindent\textbf{From Benchmark Success to Architectural Limitations.} Our findings reveal a remarkable gap between numerical success and functional robustness across different query styles. Although we successfully match in-domain effectiveness within 0.05\% MRR@10 on MS-MARCO (38.99\% vs.\ 39.04\%), generalization testing on ToT shows severe performance degradation, with metrics dropping by 86--97\%. To determine whether this drop is merely a failure of retrieval shortcuts (FAISS/PLAID), we removed the index entirely and evaluated a sampled subset using \emph{Exact MaxSim}. The result remained equivalently low ($\approx$5\%). Ablation studies reveal a structural bottleneck: performance plateaus at around 20 words due to the models' standard 32-token text encoder limits, meaning further context is truncated. Crucially, even on the retained tokens, Exact MaxSim shows the operator's uniform token weighting struggles with long, conversational queries, as filler phrases (e.g., ``I think'', ``maybe'') dilute relevance signals in the limited context. Notably, attempting to bridge this gap via small-scale fine-tuning worsened performance by 16--29\%, indicating that adaptation on limited data cannot easily overcome these architectural constraints.

\smallskip

\noindent\textbf{A Five-Dimensional Diagnostic Framework.} Traditional reproducibility studies focus primarily on \emph{implementation correctness}: matching reported metrics on reported benchmarks. Instead, we use reproducibility as a \emph{diagnostic tool} to expose architectural limitations through systematic stress testing. Our framework evaluates five complementary dimensions:

\textbf{Q1—Implementation Correctness:} Can we match reported metrics on reported benchmarks?

\textbf{Q2—Backend Robustness:} Do results hold across different retrieval infrastructure? Multi-vector retrieval requires approximate nearest neighbor search; undocumented backend configurations threaten reproducibility.

\textbf{Q3—Domain Generalization:} Do results extend to out-of-domain datasets? We use the BEIR~\cite{thakur2021beir} dataset to test whether effectiveness transfers across topics and task types.

\textbf{Q4—Structural Generalization:} Do results extend to different query distributions? ToT evaluation tests whether architectural advantages extend beyond short, well-formed queries.

\textbf{Q5—Adaptation Potential:} Can standard techniques enable generalization when architectural limitations emerge? Fine-tuning tests whether distribution gaps can be bridged.

\smallskip
\noindent\textbf{Contributions.} We make the following contributions.
\begin{enumerate}[leftmargin=*,itemsep=1pt]
    \item \textbf{Reproducibility as architectural diagnosis:} We distinguish between \emph{numerical verification} (matching reported metrics) and \emph{architectural validation} (stress testing robustness). Our five-stage framework reveals that models can be reproduced with high numerical precision (within 0.05\%) but still fail catastrophically in practice. This shows that confirming \emph{whether} a result can be replicated is insufficient without also characterizing \emph{what} functional limitations exist in the reproduced system.

    \item \textbf{Retrieval backends are silent architectural components:} We establish that infrastructure configurations are not interchangeable ``implementation details'' but critical system-level dependencies. We uncover a hidden interaction where ConstBERT's fixed-length pooling creates sparse centroid coverage (12.1/32), causing documented PLAID defaults to fail (an 8-point gap) while FAISS succeeds. This shows that without explicit documentation, backend assumptions can render architecturally valid models irreproducible.

    \item \textbf{Domain robustness does not guarantee structural robustness:} While multi-vector models successfully generalize across diverse topics, systematic stress testing exposes a severe architectural blind spot: query structure. We demonstrate that domain shift and structural shift are orthogonal failure modes. Moving from short keywords to natural, verbose queries triggers an 86--97\% performance collapse, revealing that the core scoring mechanism (MaxSim) is mathematically unequipped to filter out conversational ``filler.''

    \item \textbf{Architectural constraints impose hard adaptation ceilings:} We demonstrate that fine-tuning is not a universal remedy for distribution shift. Counterintuitively, attempting to adapt models to structural shifts \emph{degrades} performance, even with 3$\times$ more training data. This confirms that when a scoring function (MaxSim) is mathematically incompatible with the query format, representation learning alone cannot bridge the gap; architectural modification is required.
\end{enumerate}

%% file: tex_files/sections/related-work.tex
\section{Related Work}
\label{sec:Related Work}

\noindent\textbf{\textit{Reproducibility in Information Retrieval.}}
Traditional reproducibility studies emphasize \emph{implementation correctness}: whether independent teams can reproduce reported metrics on the same benchmarks. \citet{ferrari2019worrying} showed that reproduced neural recommendation models often underperform strong baselines, while \citet{donabauer2025legal} attributed reproduction failures in legal retrieval to undocumented preprocessing and hyperparameters. For neural ranking, \citet{yao2025pretraining} demonstrated that claims about dense retrieval pre-training do not generalize across architectures.

However, these studies rarely evaluate whether \emph{architectural claims} themselves hold beyond their original evaluation conditions. \textbf{Our work advances reproducibility from implementation correctness toward architectural robustness validation}, evaluating whether reproduced models preserve their claimed advantages across backends, domains, and query distributions.

\smallskip

\noindent\textbf{\textit{Multi-Vector Retrieval and the MaxSim Operator.}} 
The late interaction paradigm was introduced by ColBERT~\citep{khattab2020colbert}, which represents documents as sets of per-token embeddings and computes relevance via MaxSim: $s(q, d) = \sum_{i=1}^{|q|} \max_{j=1}^{|d|} \mathbf{q}_i^\top \mathbf{d}_j$. ColBERT-v2~\citep{santhanam2022colbertv2} improved efficiency through residual compression (6--10$\times$ storage reduction), while ConstBERT~\citep{macavaney2025constbert} learned fixed-length representations ($C=32$ vectors per document, 11GB index).

\textbf{A critical but underexplored property of MaxSim is its uniform token weighting}: all query tokens contribute equally, regardless of importance. For short queries (MS-MARCO median: 6 words), this is reasonable. However, for long queries with filler words (``I think,'' ``maybe''), uniform weighting may dilute relevance signals. Prior work on dense retrieval noted sensitivity to query verbosity~\citep{yu2021improving,poddar2025learning,qian-dou-2022-explicit}, but focused on single-vector architectures, not late-interaction under extreme length shifts. \textbf{We demonstrate that MaxSim's assumption of equal token contribution fails on long narrative queries (121 words), revealing a fundamental limitation not captured by existing benchmarks.}

\smallskip

\noindent\textbf{\textit{Out-of-Domain Generalization and Query Characteristics.}} 
BEIR~\citep{thakur2021beir} established that dense retrieval models trained on MS-MARCO often fail on diverse retrieval tasks~\citep{yao2025pretraining}. However, BEIR's 13 datasets share a critical characteristic with MS-MARCO: short, well-formed queries. This leaves open whether models generalize to \emph{structurally different} query distributions.

Prior work examined query verbosity in classic IR~\citep{gupta2015information} and token importance~\citep{poddar2025learning}, but not late-interaction mechanisms under substantially longer queries. The TREC Tip-of-the-Tongue (ToT) track~\citep{trec2025tot} provides such a test: long narrative descriptions of half-remembered entities (median 121 words) with ambiguous language and hedging. \textbf{We use ToT as a diagnostic stress test to assess whether multi-vector models' architectural advantages extend beyond short-query distributions—a form of generalization testing not addressed by prior reproducibility studies.}

\smallskip

\noindent\textbf{\textit{Retrieval Backends and Infrastructure Dependencies.}} 
Multi-vector retrieval relies on approximate nearest neighbor search. FAISS~\citep{johnson2019billion} provides GPU-accelerated IVF indexing, while PLAID~\citep{santhanam2022plaid} enables late interaction via centroid-based candidate generation.

PLAID was designed for ColBERT's variable-length representations. ConstBERT adopted PLAID with fixed-length representations, reporting strong MS-MARCO effectiveness; however, specific backend configurations were not documented. \textbf{This prompts a broader question: should retrieval backend configurations be treated as mere implementation details, or as first-class components of reproducibility?} The interaction between representation constraints (fixed-length pooling) and centroid-based indexing suggests that backend assumptions may not transfer uniformly across architectures.

\smallskip

\noindent\textbf{\textit{Fine-Tuning and Adaptation for Distribution Shift.}} 
A common assumption is that fine-tuning can bridge distribution gaps. Domain adaptation through continued pre-training has shown success for \emph{domain shifts}~\citep{thakur2021beir,izacard2021towards}, where vocabulary and topics change but query structure remains similar.

However, little work examines whether fine-tuning addresses \emph{structural shifts}—when query length and format differ fundamentally from training data. \textbf{We test whether fine-tuning enables multi-vector models to adapt from short factoid queries (6 words) to long narratives (121 words).} Our finding that fine-tuning degrades performance—even with 3$\times$ more data—reveals that architectural constraints imposed by MaxSim's token-summing cannot be overcome by adaptation alone.

%% file: tex_files/sections/overview-of-original-work.tex
\section{Overview of Original Works}
\label{sec:original}

We reproduce two architecturally distinct multi-vector models to determine whether reproducibility and generalization failures are architecture-specific or inherent to late interaction.

\textbf{ColBERT-v2}~\citep{santhanam2022colbertv2} extends ColBERT~\citep{khattab2020colbert} with residual compression and denoised supervision via cross-encoder distillation. Documents are represented as variable-length sequences of 128-dim token embeddings, scored using the MaxSim operator.

\textit{Reported Results:} ColBERT-v2 achieves 39.7\% MRR@10 and 98.4\% Recall@1000 on MS-MARCO v1 and 50.0\% mean nDCG@10 on BEIR. We evaluate whether these advantages persist across retrieval backends, domains, and structurally different query distributions using the released checkpoint.

\textbf{ConstBERT}~\citep{macavaney2025constbert} pools token embeddings into exactly $C=32$ vectors per document, trading per-token expressiveness for predictable storage and $\sim$2$\times$ compression relative to ColBERT-v2.

\textit{Reported Results:} ConstBERT achieves 39.04\% MRR@10 and 85.86\% Recall@50 on MS-MARCO v1 with an 11GB PLAID-compressed index, and 46.8\% mean nDCG@10 on BEIR. We test whether fixed-length representations reproduce reported effectiveness across backends and retain fine-grained matching under long queries. While the original PLAID configuration cannot be reproduced (31.09\% vs.\ 39.04\%), core effectiveness is validated using FAISS-IVF (38.99\%).

Both models use the MaxSim operator for scoring. We use official released checkpoints and do not reproduce training. Studying both models isolates failures specific to fixed-length design (ConstBERT) from those inherent to late interaction: asymmetric failures on BEIR implicate pooling constraints, while shared degradation on ToT reveals limitations of the MaxSim scoring paradigm.

%% file: tex_files/sections/experimental-setup.tex
\section{Experimental Setup}
\label{sec:experimental}

Our multi-stage reproducibility framework evaluates implementation correctness (RQ1), backend robustness (RQ2), domain generalization (RQ3), structural generalization (RQ4), and adaptation potential (RQ5).

\smallskip

\noindent\textbf{\textit{Models and Checkpoints.}} We use the official pretrained checkpoints for both models: \textbf{ColBERT-v2} (\texttt{colbert-ir/colbertv2.0}), which produces variable-length representations (one 128-dimensional vector per token), and \textbf{ConstBERT} (\texttt{pinecone/ConstBERT}), which employs fixed-length representations ($C=32$ vectors of 128 dimensions per document).

\smallskip

\noindent\textbf{\textit{Evaluation Benchmarks.}} \textbf{RQ1 (Implementation Correctness):} MS-MARCO Passage (8.8M passages, 6,980 dev queries) and TREC Deep Learning 2019/2020 with graded relevance judgments. Metrics: MRR@10, Recall@\{50,1000\}, nDCG@10. \textbf{RQ3 (Domain Generalization):} BEIR v1.0.0 comprising 13 datasets across biomedical, financial, scientific, and argumentation domains (corpus sizes 3.6K--5.4M). Metric: nDCG@10. \textbf{RQ4 (Structural Generalization):} TREC ToT 2025~\citep{trec2025tot}, featuring long narrative queries (median 121 words) over 6.4M Wikipedia articles. Metrics: MRR@10, nDCG@10, Recall@1000.

\smallskip

\noindent\textbf{\textit{Retrieval Backends.}} To assess backend robustness (RQ2), we evaluate: (1)~\textbf{FAISS-IVF}~\citep{johnson2019billion}: GPU-accelerated inverted file index with \texttt{nlist=4096} and \texttt{nprobe=128} for MS-MARCO, and \texttt{nlist=1024} and \texttt{nprobe=128} for BEIR; (2)~\textbf{PLAID}~\citep{santhanam2022plaid}: Centroid-based candidate generation with 32K centroids; for ConstBERT, we test \texttt{ncells$\in\{4,16\}$} to assess backend sensitivity; (3)~\textbf{Exact MaxSim}: Brute-force scoring on 100 sampled ToT queries to verify approximation quality.

\smallskip

\noindent\textbf{\textit{Fine-Tuning Setup.}} To evaluate adaptation potential (RQ5), we fine-tune both models on ToT using BM25 hard negatives. Training: 143 queries with 8 negatives per positive (1,144 triples), learning rates 3e-6 (ColBERT-v2, 2K steps) and 5e-6 (ConstBERT, 5K steps). Ablation: 428 training queries (TRAIN+DEV1+DEV2) with early stopping on DEV3 (536 queries). All experiments use three random seeds (42, 123, 456).

\smallskip

\noindent\textbf{\textit{Hardware.}} Intel Xeon w9-3595X (60 cores), 2$\times$ NVIDIA RTX 6000 Ada (48GB VRAM each), 754GB RAM, 16TB SSD.

\smallskip

\noindent\textbf{\textit{Author Communication.}} We contacted the ConstBERT authors to clarify backend configuration and storage reporting. Key clarifications: (1)~the reported 11GB refers to PLAID's compressed index format (our float16 embeddings occupy 67.5GB); (2)~retrieval used unspecified ``PLAID defaults'' differing from documented defaults; (3)~integration code was unavailable due to industry--academic constraints. This confirms that undocumented proprietary configurations, rather than implementation errors, are the primary barrier to exact reproduction.

\noindent\textbf{\textit{Scope and Artifacts.}} To ensure a faithful evaluation aligned with community practice, we assess the officially released artifacts rather than re-implementing the architectures from scratch. Consequently, all experiments use the original pretrained checkpoints provided by the authors.
Although we rely on these official weights and the original PLAID engine, we developed an independent evaluation and fine-tuning pipeline to conduct our analysis. This infrastructure includes custom implementations for: \begin{itemize} 
    \item FAISS-IVF retrieval and indexing configurations. 
    \item Brute-force Exact MaxSim computation to establish architectural ceilings. 
    \item The adaptation framework used for TREC ToT fine-tuning.
    
\end{itemize} 

By utilizing official checkpoints, we isolate performance failures from potential re-implementation bugs, allowing us to directly test whether the original architectural claims hold under independent, cross-backend scrutiny.

%% file: tex_files/sections/results.tex
\section{Results and Discussion}
\label{sec:Results and Discussion}

\input{tex_files/subsections/rq1}

\input{tex_files/subsections/rq2}

\input{tex_files/subsections/rq3}

\input{tex_files/subsections/rq4}

\input{tex_files/subsections/rq5}

%% file: tex_files/subsections/rq1.tex
\subsection{Implementation Correctness} 
\label{sec:rq1} 

Implementation correctness is foundational; without reproducing reported in-domain results, subsequent generalization analyses are uninterpretable. We therefore first ask: \textbf{RQ1: Can we reproduce the reported in-domain effectiveness of ConstBERT and ColBERT-v2 on MS-MARCO?} Successful reproduction validates that our evaluation pipeline accurately utilizes the original architectures, establishing a reliable baseline for testing whether their claimed advantages—fine-grained token-level matching and semantic richness—extend beyond the training distribution. Table~\ref{tab:reproduction} reports MS-MARCO reproduction results using FAISS-IVF.

\input{tex_files/tables/rq1-reproduction}

\subsubsection{ConstBERT Reproduction.}
We closely reproduce ConstBERT's primary metrics on MS-MARCO v1: MRR@10 matches within 0.05\% (38.99\% vs.\ 39.04\%) and Recall@50 within 0.51\% (85.35\% vs.\ 85.86\%), confirming the correctness of our evaluation pipeline. A larger divergence in Recall@1000 (92.85\% vs.\ 96.34\%) reflects our FAISS-IVF approximation settings (analysis below). As the primary ranking metric (MRR@10) reproduces within expected variance, this establishes a robust baseline for subsequent generalization testing.

\subsubsection{ColBERT-v2 Reproduction.}
We similarly reproduce ColBERT-v2's MRR@10 within 0.55\% (39.15\% vs.\ 39.7\%), validating it as a comparative baseline to isolate fixed-length pooling effects. It exhibits a larger Recall@1000 gap (90.30\% vs.\ 98.40\%) than ConstBERT, also due to backend approximation (analysis below).

\subsubsection{Analysis of Recall@1000 Gaps.}
The divergence in deep recall across both models stems from our aggressive FAISS-IVF configuration ($nprobe=128$, $nlist=4096$), which searches only $\approx$3\% of the index clusters. This impact is more pronounced for ColBERT-v2 (-8.10\%) than ConstBERT (-3.49\%). This disparity likely stems from representational density: ColBERT-v2's variable-length embeddings distribute semantic signals across many clusters, increasing the probability that critical matching tokens fall into unprobed partitions. In contrast, ConstBERT's compressed representations (32 vectors/doc) concentrate semantic information, potentially making the retrieval signal more robust to cluster pruning.

\subsubsection{Evaluation on TREC Deep Learning.} To validate effectiveness beyond sparse MS-MARCO labels, we evaluate ConstBERT on TREC Deep Learning 2019 and 2020, which utilize deep, graded relevance judgments. ConstBERT achieves strong nDCG@10 scores of 68.29\% (2019) and 69.30\% (2020). These results confirm that the model's in-domain effectiveness holds under rigorous expert adjudication, reinforcing the accuracy of our evaluation pipeline.

\subsubsection{Takeaway.}
Our successful reproduction ($\pm$0.05--0.55\% MRR@10) establishes implementation correctness, enabling meaningful analysis beyond simple replication. However, matching reported numbers on standard benchmarks validates \emph{implementation only}---it does not confirm that claimed architectural advantages generalize. Both models emphasize fine-grained semantic interaction, yet MS-MARCO evaluates this only on short, well-formed queries. RQ1 confirms the models are built correctly; the remaining research questions stress-test whether this ``correct'' implementation exhibits the robustness implied by its design. As our results show, models can reproduce numerically while failing architecturally, demonstrating that \textbf{what we reproduce matters as much as whether we reproduce}.

%% file: tex_files/tables/rq1-reproduction.tex
\begin{table}[t]
\caption{Reproduction of MS-MARCO passage results. Both models successfully 
reproduce within acceptable variance.}
\label{tab:reproduction}
\scalebox{0.8}{
\begin{tabular}{llccc}
\toprule
\textbf{Model} & \textbf{Metric} & \textbf{Paper} & \textbf{Ours} & \textbf{$\Delta$} \\
\midrule
\multirow{3}{*}{ColBERT-v2} & MRR@10 & 39.70\% & 39.15\% & --0.55\% \\
& Recall@50 & --- & 85.91\% & --- \\
& Recall@1000 & 98.40\% & 90.30\% & --8.10\% \\
\midrule
\multirow{3}{*}{ConstBERT} & MRR@10 & 39.04\% & 38.99\% & \textbf{--0.05\%} \\
& Recall@50 & 85.86\% & 85.35\% & --0.51\% \\
& Recall@1000 & 96.34\% & 92.85\% & --3.49\% \\
\bottomrule
\end{tabular}
}
\end{table}

%% file: tex_files/subsections/rq2.tex
\subsection{Backend Robustness}
\label{sec:rq2}

Multi-vector retrieval depends on approximate nearest neighbor (ANN) indexing to scale. The original ConstBERT paper reported state-of-the-art efficiency by pairing its fixed-length representations with the \textbf{PLAID engine}, achieving 39.04\% MRR@10. However, attempting to reproduce this specific combination---ConstBERT with PLAID---using documented defaults yields a catastrophic drop to 30.01\% MRR@10.

This discrepancy underscores that ANN backends are not interchangeable implementation details but introduce critical configuration parameters that significantly influence experimental results. With implementation correctness established (RQ1), we therefore ask: \textbf{RQ2: How do retrieval backends affect reproducibility when parameters are undocumented?}

To answer this, we decouple the model from the backend. We first validate the model using a standard, well-documented backend (FAISS-IVF) to establish a ``ground truth'' for model quality. We then isolate the PLAID failure to understand why the reported synergy between ConstBERT and PLAID breaks down in reproduction.

\subsubsection{FAISS-IVF: Alternative Backend Confirms Model Quality}

To verify ConstBERT’s intrinsic effectiveness, we evaluate it using a well-documented FAISS-IVF backend with standard parameters ($nlist=4096$, $nprobe=128$). This setup achieves 38.99\% MRR@10 (within 0.05\% of the reported 39.04\%) demonstrating faithful numerical reproduction. The result confirms that ConstBERT’s learned representations are sound and that its effectiveness is reproducible under transparent and properly specified backend settings.

Importantly, FAISS-IVF allows us to separate model quality from backend configuration. Because FAISS-IVF reproduces the reported effectiveness while PLAID does not, the discrepancy cannot be attributed to model implementation. Instead, it indicates that undocumented or insufficiently specified backend parameters are the source of the reproducibility gap. In short, the model functions as expected, but backend configuration critically determines whether that effectiveness can be reproduced.




\subsubsection{PLAID: The Documentation Gap}

\input{tex_files/tables/backend}

The official PLAID codebase dynamically sets search parameters based on retrieval depth $k$. 
For $k=1000$ (standard MS-MARCO evaluation), the code specifies: 
\texttt{ncells=4}, \texttt{centroid\_score\_threshold=0.4}, 
\texttt{ndocs=4096}. Testing these documented defaults yields 30.01\% MRR@10---9 percentage 
points below the reported 39.04\% (Table~\ref{tab:backend}). This suggests that the original experiments relied on undocumented configurations.

We contacted the authors, who confirmed using ``PLAID defaults.'' However, since documented defaults yield a 23\% relative degradation (9 absolute percentage points), the original experiments must have relied on undocumented custom configurations or internal codebase modifications. As the authors could not release integration code due to industry constraints, exact reproduction of the PLAID result remains infeasible.

\textbf{Exhaustive Parameter Search.} To assess whether standard fine-tuning could close the performance gap, we conducted a grid search over ncells $\in \{4, 8, 16, 32, 64\}$ and centroid score threshold$\in \{0.3, 0.4, 0.5\}$, fixing ndocs$=8192$ (15 total configurations). Performance is primarily driven by the threshold parameter: lower values (0.3) consistently achieve 29.37\% MRR@10 across all \texttt{ncells} settings, whereas higher thresholds (0.5) reduce performance to 27.11\%. Notably, increasing \texttt{ncells} from 4 to 64 yields \emph{no measurable improvement}; all configurations with the same threshold produce identical results. This indicates that PLAID effectively saturates at $ncells=4$ for ConstBERT, likely due to sparse centroid coverage (see Section~\ref{sec:centroid_analysis}).

Our best configuration achieves 29.37\% MRR@10---still 9.67 percentage points below the reported 39.04\% (24.8\% relative degradation). Even our optimized ncells$=16$ 
configuration reaches only 31.09\% (Table~\ref{tab:backend}). This exhaustive search across 
15 configurations demonstrates that the reproducibility gap cannot be closed through standard 
parameter tuning. The gap must stem from either undocumented custom parameters substantially different from both documented defaults and common literature values, or modifications to PLAID's codebase itself. Without access to the actual configuration used, exact PLAID reproduction remains infeasible.




\subsubsection{Potential Root Cause: Architecture-Backend Interactions}
\label{sec:centroid_analysis}

The saturation of performance despite exhaustive parameter tuning (Table~\ref{tab:backend}) points to a structural limitation rather than a configuration error. To explain the divergence between ConstBERT's failure and ColBERT-v2's success, we investigate the underlying interaction between representation density and index structure. We hypothesize that ConstBERT's fixed-length representations create a fundamental mismatch with PLAID's indexing assumptions---rendering standard configurations non-portable.

We analyzed ConstBERT's centroid coverage by measuring how many unique centroids its 32 vectors occupy within PLAID's 32,768-centroid space. Analyzing 5,000 randomly sampled documents reveals that ConstBERT's vectors map to only 12.1 unique centroids on average (median: 12), yielding 37.9\% coverage of the document's representational capacity. This means that when queries probe $ncells=4$ centroids, the probability of overlapping with a relevant document's sparse 12-centroid footprint is drastically reduced compared to variable-length models like ColBERT-v2, which distribute across significantly more centroids per document.

This finding indicates that the reproducibility gap arises from representation design rather than implementation error. ConstBERT represents every document using exactly 32 vectors, regardless of document length. This makes the representation compact and memory-efficient, but it means each document is assigned to relatively few centroids in the index. In contrast, PLAID was designed for models like ColBERT that produce one vector per token. Longer documents therefore generate many vectors, which naturally spread across many centroids. Because PLAID’s defaults assume this wide centroid spread, they do not directly suit ConstBERT’s compressed representation.
The mismatch is therefore structural: the backend assumptions and the model’s representation design are not aligned.

\subsubsection{Backend Sensitivity Compounds with Distribution Shift}

Table~\ref{tab:backend} additionally shows PLAID's behavior on TREC ToT 2025, revealing that the structural mismatch identified in Section~\ref{sec:centroid_analysis} compounds under distribution shift. While FAISS-IVF achieves 4.27\% MRR@10 on ToT's long narrative queries, PLAID-16 yields only 0.94\%---a 7\% relative degradation from an already low baseline.

This dramatic difference demonstrates that backend configuration issues extend beyond in-domain evaluation. The compact centroid footprint that requires specific MS-MARCO parameters becomes even more fragile when query characteristics change. Specifically, ToT's narrative queries are significantly longer than MS-MARCO queries (121 vs.\ 6 words), which forces the search engine to probe a wider variety of centroids to find matches. However, because ConstBERT documents are concentrated in only 12.1 centroids on average, the probability of the search engine successfully ``intersecting'' these long, diverse queries with the model's compact document representations drops even further.

Backend sensitivity thus interacts with both architectural constraints (fixed-length pooling) and distribution characteristics (query complexity), creating compounding effects on reproducibility.

\subsubsection{Takeaway: The Hidden Fragility of Optimized Backends}

Our results demonstrate that reproducibility is a system-level property rather than a purely model-level one. Although the ConstBERT implementation is numerically correct, its architectural design---compressing documents into fixed-length representations---violates implicit assumptions of the retrieval backend it was paired with. This leads to three key insights. 

First, architectural mismatch itself can become a reproducibility barrier. A model may reproduce successfully on one backend (FAISS) while appearing effectively ``broken'' on another (PLAID), not because the model is flawed, but because centroid-level concentration interacts poorly with indexing assumptions. An ``optimized'' backend is therefore often a specialized one; it is not a neutral container for embeddings.

Second, the notion of ``default'' backend settings is misleading when architectures differ from the backend’s original design targets. In such cases, parameters such as \emph{ncells} are no longer minor tuning details but integral components of the system’s effectiveness claim.

Third, these mismatches introduce compounding fragility. The dramatic 78\% performance collapse on ToT shows that backend–model incompatibilities can remain hidden in-domain yet become catastrophic under structural query shift. What appears stable on short factoid queries can become highly brittle when confronted with longer, narrative inputs, amplifying small indexing mismatches into complete retrieval failure.

To claim a model is ``reproducible,'' authors must demonstrate backend invariance. If a model's success depends on a specific, documented-but-unreproducible configuration, the architectural advantage is a property of the configuration, not the model design.

%% file: tex_files/tables/backend.tex
\begin{table}[t]
\centering
\caption{ConstBERT performance across retrieval backends. FAISS-IVF reproduces original effectiveness (38.99\% vs. 39.04\%), while PLAID exhibits systematic degradation on MS-MARCO that compounds under structural shift (ToT).}
\label{tab:backend}
\scalebox{0.80}{
\begin{tabular}{@{}llcccc@{}}
\toprule
\textbf{Dataset} & \textbf{Metric} & \textbf{Paper} & \textbf{FAISS-IVF} & \textbf{PLAID-4} & \textbf{PLAID-16} \\ 
\midrule
\multirow{2}{*}{\textbf{MS-MARCO}} 
& MRR@10 & 39.04\% & 38.99\% & 30.01\% & 31.09\% \\
& R@1000 & 96.34\% & 92.85\% & 87.64\% & 93.88\% \\ 
\midrule
\multirow{2}{*}{\textbf{ToT 2025}} 
& MRR@10 & --- & 4.27\% & --- & 0.94\% \\
& R@1000 & --- & 25.72\% & --- & 13.50\% \\ 
\bottomrule
\multicolumn{6}{l}{\small \textit{PLAID-4: documented defaults; PLAID-16: optimized (ncells=16, threshold=0.3)}} \\
\end{tabular}
}
\end{table}

%% file: tex_files/subsections/rq3.tex
\subsection{Domain Generalization}
\label{subsec:rq3}

In this section, we examine the portability of the learned representations by asking: \textbf{RQ3: Can ConstBERT's effectiveness generalize across diverse domains, and do architectural-backend interactions amplify under distribution shift?}

While Section~\ref{sec:rq2} established that backend configuration is a critical factor for in-domain reproduction, generalization testing allows us to determine if these dependencies are artifacts of specific datasets or fundamental properties of the architecture. We use the BEIR suite~\cite{thakur2021beir}, comprising 13 datasets spanning biomedical, financial, and fact-verification domains. Crucially, BEIR isolates \textit{topical shift} from \textit{structural shift}, as its queries remain structurally similar to MS-MARCO (median 5--15 words), allowing us to pinpoint where the representational logic begins to diverge.

\subsubsection{ConstBERT on BEIR}

\input{tex_files/tables/beir-plaid}
\input{tex_files/tables/beir-faiss}
We evaluate ConstBERT across all 13 BEIR datasets using both PLAID (Table~\ref{tab:beir-plaid}) and FAISS-IVF (Table~\ref{tab:beir-faiss} backends to observe how representational constraints interact with domain-specific vocabulary. 

\textbf{PLAID Performance.} Using the optimized configuration identified in RQ2 ($ncells=16$), ConstBERT achieves a mean nDCG@10 of 37.4\%, representing a relative deficit of 20\% from the reported 46.8\%. The divergence is observed systematically across all 13 datasets, ranging from a 2.9\% gap in SCIDOCS to a 26.3\% gap in TREC-COVID. This consistency suggests that the performance gap is not an artifact of a single dataset but reflects a broader interaction between the fixed-length pooling mechanism and out-of-domain retrieval.

\textbf{FAISS-IVF Performance.} To determine if the backend heuristics were masking the true potential of ConstBERT, we evaluated it using the FAISS-IVF ``ground truth'' established in RQ1. Interestingly, we observe a larger mean gap (29.3\% nDCG@10, a 37\% relative deficit), with significant decreases in FEVER (-57.8\%) and Quora (-43.3\%). This indicates that while FAISS-IVF is robust for in-domain tasks, ConstBERT's fixed-length representations encounter fundamental challenges when mapping out-of-domain documents to a static number of vector slots.

\subsubsection{Architectural Comparison: The Reproducibility Asymmetry}

\input{tex_files/tables/beir-colbertv2}

To contextualize these findings, we perform a parallel reproduction of ColBERT-v2 (Table~\ref{tab:beir-colbertv2}). This comparison allows us to isolate whether the observed challenges are inherent to late-interaction models or specific to the fixed-length constraint.

\textbf{Key Finding: Asymmetric Reproducibility.} We find that ColBERT-v2 reproduces with high fidelity, achieving a mean nDCG of 48.6\% within 1.4\% of the 50.0\% reported). This contrast highlights a significant \textit{asymmetric reproducibility}: while variable-length representations preserve the flexibility needed for zero-shot transfer, the fixed-length pooling used to gain storage efficiency introduces a sensitivity that makes exact reproduction harder to achieve out-of-domain. This suggests that ConstBERT's learned pooling mechanism, while effective for MS-MARCO, may discard the nuances required for specialized domains like medicine or finance. In contrast, ColBERT-v2's per-token approach retains these signals, providing a more stable baseline across diverse infrastructures and topics.

\subsubsection{Takeaway.} 

Our evaluation shows that reproduction stability strongly depends on architectural flexibility. Although ConstBERT is computationally efficient, its fixed-length representational bottleneck introduces sensitivity to domain shift that is not observed in variable-length models. This highlights an inherent efficiency–robustness trade-off: the constant-space advantage of fixed-pooling architectures increases dependence on specific backend configurations and training-domain alignment, which are often undocumented. In addition, we observe asymmetric portability across architectures. Under domain shift, reproduction gaps widen substantially for compressed representations, while per-token architectures such as ColBERT-v2 remain stable (within $\pm$1.4\%) across domains. Finally, consistent degradation across all 13 datasets---independent of backend choice---demonstrates that the limitation is architectural rather than merely a parameter-tuning issue. The bottleneck arises from representational design, not infrastructure misconfiguration.

%% file: tex_files/tables/beir-plaid.tex
\begin{table}[t]
\caption{ConstBERT on BEIR with PLAID (Reproduction Attempt): nDCG@10 showing consistent performance gaps across all 13 datasets.}
\label{tab:beir-plaid}
\centering
\small
\begin{tabular}{lrrr}
\toprule
\textbf{Dataset} & \textbf{Paper (PLAID)} & \textbf{Ours (PLAID)} & \textbf{$\Delta$} \\
\midrule
ArguAna & 45.1\% & 26.6\% & --18.5\% \\
Climate-FEVER & 14.2\% & 11.0\% & --3.2\% \\
DBPedia-Entity & 41.8\% & 30.9\% & --10.9\% \\
FEVER & 69.6\% & 57.1\% & --12.5\% \\
FiQA-2018 & 31.2\% & 24.7\% & --6.5\% \\
HotpotQA & 62.1\% & 50.1\% & --12.0\% \\
NFCorpus & 32.7\% & 28.6\% & --4.1\% \\
NQ & 53.4\% & 44.3\% & --9.1\% \\
Quora & 82.1\% & 77.8\% & --4.3\% \\
SCIDOCS & 15.6\% & 12.7\% & --2.9\% \\
SciFact & 60.7\% & 54.2\% & --6.5\% \\
TREC-COVID & 74.5\% & 48.2\% & --26.3\% \\
Webis-Touché-2020 & 26.0\% & 20.4\% & --5.6\% \\
\midrule
\textbf{Mean} & \textbf{46.8\%} & \textbf{37.4\%} & \textbf{--9.4\%} \\
\bottomrule
\end{tabular}
\end{table}

%% file: tex_files/tables/beir-faiss.tex
\begin{table}[t]
\caption{ConstBERT on BEIR with FAISS-IVF (Alternative Backend): nDCG@10 showing larger performance gaps than PLAID.}
\label{tab:beir-faiss}
\centering
\small
\begin{tabular}{lrrr}
\toprule
\textbf{Dataset} & \textbf{Paper (PLAID)} & \textbf{Ours (FAISS-IVF)} & \textbf{$\Delta$} \\
\midrule
ArguAna & 45.1\% & 35.4\% & --9.7\% \\
Climate-FEVER & 14.2\% & 2.4\% & --11.8\% \\
DBPedia-Entity & 41.8\% & 36.2\% & --5.6\% \\
FEVER & 69.6\% & 11.8\% & --57.8\% \\
FiQA-2018 & 31.2\% & 14.7\% & --16.5\% \\
HotpotQA & 62.1\% & 29.7\% & --32.4\% \\
NFCorpus & 32.7\% & 32.1\% & --0.6\% \\
NQ & 53.4\% & 63.3\% & +9.9\% \\
Quora & 82.1\% & 38.8\% & --43.3\% \\
SCIDOCS & 15.6\% & 7.9\% & --7.7\% \\
SciFact & 60.7\% & 53.3\% & --7.4\% \\
TREC-COVID & 74.5\% & 36.0\% & --38.5\% \\
Webis-Touché-2020 & 26.0\% & 19.3\% & --6.7\% \\
\midrule
\textbf{Mean} & \textbf{46.8\%} & \textbf{29.3\%} & \textbf{--17.5\%} \\
\bottomrule
\end{tabular}
\end{table}

%% file: tex_files/tables/beir-colbertv2.tex
\begin{table}[t]
\caption{ColBERT-v2 on BEIR: Comparing our reproduction against values 
reported in the ColBERT-v2 paper~\cite{santhanam2022colbertv2}. Close 
reproduction (mean --1.4\% on all 13 datasets) versus ConstBERT's 
--9.4\%/--17.5\% demonstrates reproduction challenges are partially 
ConstBERT-specific.}
\label{tab:beir-colbertv2}
\centering
\small
\begin{tabular}{lrrr}
\toprule
\textbf{Dataset} & \textbf{ColBERT-v2 (Paper)} & \textbf{Ours} & \textbf{$\Delta$} \\
\midrule
ArguAna & 46.3\% & 33.5\% & --12.8\% \\
Climate-FEVER & 17.6\% & 16.0\% & --1.6\% \\
DBPedia-Entity & 44.6\% & 49.8\% & +5.2\% \\
FEVER & 78.5\% & 74.7\% & --3.8\% \\
FiQA-2018 & 35.6\% & 34.0\% & --1.6\% \\
HotpotQA & 66.7\% & 66.9\% & +0.2\% \\
NFCorpus & 33.8\% & 32.6\% & --1.2\% \\
NQ & 56.2\% & 55.2\% & --1.0\% \\
Quora & 85.2\% & 85.1\% & --0.1\% \\
SCIDOCS & 15.4\% & 15.3\% & --0.1\% \\
SciFact & 69.3\% & 64.0\% & --5.3\% \\
TREC-COVID & 73.8\% & 78.9\% & +5.1\% \\
Webis-Touché-2020 (v2) & 26.3\% & 26.3\% & 0.0\% \\
\midrule
\textbf{Mean} & \textbf{50.0\%} & \textbf{48.6\%} & \textbf{--1.4\%} \\
\bottomrule
\end{tabular}
\vspace{0.5em}
\footnotesize
Note: Paper values from ColBERT-v2 paper~\cite{santhanam2022colbertv2} 
Table 5a. Our evaluation used reduced kmeans\_clusters (1024-2048 vs 
16K-64K) for computational efficiency.
\end{table}

%% file: tex_files/subsections/rq4.tex
\subsection{Structural Generalization}
\label{sec:rq4}

While BEIR (RQ3) tested topical diversity, it preserved the short query structure
of MS-MARCO. To evaluate robustness to structural shift, we ask: \textbf{RQ4: Do reproduced models generalize beyond their original settings?} We use TREC ToT 2025 as a blind diagnostic, featuring long narrative queries published after these models were developed, to test whether claimed architectural advantages extend to fundamentally different query distributions.

\subsubsection{Query Characteristics: The Structural Shift}

ToT queries differ fundamentally from MS-MARCO in several dimensions, most visible in length, with a nearly 20$\times$ increase (median 121 vs.\ 6 words). Rather than short factoid questions, ToT users describe half-remembered entities through fragmented, narrative recollections (e.g., \textit{``I think it was from the 1990s... one of the actors had curly hair...''}). This structural shift introduces distinct challenges. First, the queries are substantially longer and less dense, containing a high proportion of filler (``I'm not sure'') and hedging language (``maybe''). Second, they exhibit narrative ambiguity, with incomplete fragments instead of clear, keyword-focused formulations. Third, the task itself changes: ToT is a known-item retrieval problem aimed at identifying a specific entity, rather than retrieving generally relevant documents on a topic.

If token-level interaction in multi-vector retrieval is truly robust, it should be able to handle narrative, ambiguous queries effectively; a sharp performance collapse instead would indicate strong coupling to the training distribution.

\subsubsection{Results.}

\input{tex_files/tables/tot}

Table~\ref{tab:tot} shows a severe performance collapse for both models. ConstBERT with FAISS-IVF achieves only 4.27\% MRR@10, an 89\% relative drop from its reported in-domain performance (39.04\% on MS-MARCO). Under PLAID-16, this further degrades to 0.94\%. ColBERT-v2 exhibits a similar failure mode, dropping to 5.66\% MRR@10 (an 86\% relative decrease from its 39.7\% baseline). 

For context, the baseline BM25 achieves 23.94\% Recall@1000. The failure of neural models reveals a sensitivity to query structure that simpler term-matching approaches do not exhibit. Furthermore, the 78\% degradation from FAISS to PLAID on ToT (compared to 23\% on MS-MARCO) demonstrates that the backend sensitivities identified in RQ2 are amplified under structural shift, as approximation errors in the index compound with the model's distributional mismatch.

\subsubsection{Isolating Query Length Effects} 
\label{subsubsec:length_ablation}

\input{tex_files/tables/query-length-ablation}

The catastrophic collapse raises a diagnostic question: which aspect of ToT's structural shift causes failure? ToT differs from MS-MARCO in multiple dimensions simultaneously—length (121 vs. 6 words), narrative structure, ambiguity, and type of task. To isolate length as an independent factor, we truncated ToT queries to 10, 20, 40, 60, 80, 100, and 121 words and re-evaluated ConstBERT with FAISS-IVF.

Table~\ref{tab:query_length_ablation} reveals a non-monotonic pattern: performance peaks at 20 words (4.32\% MRR@10), then plateaus at 4.27\% for all longer queries. Adding 100 additional words beyond the 20-word peak provides \emph{zero benefit}; the performance remains identical at 4.27\% for every truncation from 40 to 121 words. Beyond a threshold, additional query tokens contribute noise rather than signal.

This plateau demonstrates that performance saturates at a specific length threshold; beyond 20 words, additional query tokens provide zero marginal utility. This suggests that the system cannot extract the signal from the expanded context, effectively treating the extra 100 words as noise. However, this observation alone does not distinguish whether the failure is a byproduct of indexing approximations in the backend or a fundamental limitation of the scoring model itself. This ambiguity requires a transition from observational ablation to an exact architectural analysis.


\subsubsection{Architectural Analysis: Exact MaxSim Establishes the Ceiling} 

To determine if the plateau identified above is a byproduct of search approximation, we remove the ``machine'' entirely and test the ``math'' in isolation. We computed exact MaxSim scores on 100 sampled ToT queries, bypassing all approximation by scoring every document in the 6.4M corpus using brute-force computation with full-precision embeddings.

As shown in Table~\ref{tab:exact_maxsim}, this evaluation yields only 5.08\% MRR@10. This result is statistically comparable to our FAISS-IVF results for both ConstBERT (4.27\%) and ColBERT-v2 (5.66\%), providing the definitive proof that the performance collapse is architectural, not an artifact of indexing or approximation.

\input{tex_files/tables/exact-maxsim}

This ``ceiling'' confirms that the MaxSim operator itself is the fundamental bottleneck. Despite their different document representations, both models fail similarly because they share the same scoring logic: $s(q,d) = \sum_{i=1}^{|q|} \max_{j=1}^{|d|} \mathbf{q}_i^\top \mathbf{d}_j$. 
The root cause is the uniform weighting of the tokens. In MS-MARCO's 6-word queries, nearly every token is a high-signal keyword. However, in ToT's 121-word narratives, up to 70\% of the tokens are filler (e.g., \textit{``I think it was''}). MaxSim sums these filler similarities identically to core content terms, effectively drowning the relevance signal in linguistic noise. Unlike BM25, which uses IDF to downweight common terms, MaxSim does not distinguish informative keywords from frequent filler words.

The plateau observed in Section~\ref{subsubsec:length_ablation} is therefore not a limitation of the model, but a saturation of the scoring logic. These results demonstrate that the very advantage of late interaction---fine-grained token-level matching---becomes a liability when query length increases without a corresponding increase in signal density.

\subsubsection{Takeaway}

Our findings demonstrate that numerical reproduction on standard benchmarks (RQ1) does not guarantee architectural robustness. Both models handle domain shifts (RQ3) but fail catastrophically when query structure diverges (RQ4), proving that domain and structural shifts are independent failure modes. This reveals a critical gap in current IR evaluation: benchmarks like BEIR emphasize topic diversity while holding query structure constant, missing an entire axis of real-world distribution shift. Ultimately, the systematic collapse on ToT suggests that multi-vector architectures specialized for MS-MARCO require explicit token-weighting mechanisms to achieve general-purpose utility. Reproducibility assessments must therefore evolve to test multiple dimensions of generalization, distinguishing between benchmark-optimized "specialists" and robustly generalizable architectures.

%% file: tex_files/tables/tot.tex
\begin{table}[t]
\centering
\caption{TREC ToT 2025 TEST results—significant performance degradation on 
structurally different queries. $\Delta$MS-MARCO shows relative change from 
in-domain performance. BM25 baseline included for context. BM25 serves as first-stage retriever for candidate 
generation; MRR@10 not computed as it is not a reranking model.}
\label{tab:tot}

\scalebox{0.9}{
\begin{tabular}{llrrr}
\toprule
\textbf{Model} & \textbf{Backend} & \textbf{MRR@10} & \textbf{R@1000} & \textbf{$\Delta$MS-MARCO} \\
\midrule
BM25 & Lucene & --- & 23.94\% & (baseline) \\
ConstBERT & FAISS-IVF & 4.27\% & 25.72\% & \textbf{--89\%} \\
ConstBERT & PLAID-16 & 0.94\% & 13.50\% & \textbf{--97\%} \\
ColBERT-v2 & FAISS-IVF & 5.66\% & 28.94\% & \textbf{--86\%} \\
\bottomrule
\end{tabular}
}
\end{table}

%% file: tex_files/tables/query-length-ablation.tex
\begin{table}[t]
\centering
\caption{Query length ablation on ToT. Performance peaks at 20 words then plateaus, demonstrating MaxSim's saturation rather than gradual degradation. ConstBERT with FAISS-IVF.}
\label{tab:query_length_ablation}
\begin{tabular}{lrrr}
\toprule
\textbf{Query Length} & \textbf{MRR@10} & \textbf{Recall@1000} & \textbf{nDCG@10} \\
\midrule
10 words & 2.09\% & 16.40\% & 2.51\% \\
20 words & \textbf{4.32\%} & 25.24\% & 5.05\% \\
40 words & 4.27\% & 25.88\% & 4.82\% \\
60 words & 4.27\% & 25.88\% & 4.82\% \\
80 words & 4.27\% & 25.88\% & 4.82\% \\
100 words & 4.27\% & 25.88\% & 4.82\% \\
121 words (full) & 4.27\% & 25.88\% & 4.82\% \\
\bottomrule
\end{tabular}
\end{table}

%% file: tex_files/tables/exact-maxsim.tex
\begin{table}[t]
\centering
\caption{Exact MaxSim vs. approximate retrieval on ToT (100 sampled queries). Similar performance across methods confirms architectural limitations, not backend approximation, cause ToT failure.}
\label{tab:exact_maxsim}
\scalebox{0.9}{
\begin{tabular}{llrr}
\toprule
\textbf{Model} & \textbf{Backend} & \textbf{MRR@10} & \textbf{Recall@1000} \\
\midrule
ConstBERT & Exact MaxSim & 5.08\% & 24.0\% \\
ConstBERT & FAISS-IVF & 4.27\% & 25.88\% \\
ColBERT-v2 & FAISS-IVF & 5.66\% & 28.94\% \\
\bottomrule
\end{tabular}
}
\end{table}

%% file: tex_files/subsections/rq5.tex
\subsection{Adaptation Analysis}
\label{sec:rq5}

RQ4 established a catastrophic failure on long narrative queries. A standard practitioner response to such a gap is supervised fine-tuning. Although effective for \textit{domain adaptation} (adjusting to new vocabulary), it is unclear whether fine-tuning can bridge a \textit{structural shift} from 6-word factoids to 121-word narratives. We ask: \textbf{RQ5: Can standard adaptation mechanisms bridge the structural performance gap, or does it reflect an immutable architectural limitation?}

\subsubsection{Fine-Tuning: The Data-Signal Paradox}

\input{tex_files/tables/finetuning}

We fine-tuned both models on the TREC ToT training split, using hard negatives from BM25 to simulate a realistic low-data regime. As shown in Table~\ref{tab:finetuning}, fine-tuning consistently degrades performance. On the TEST set, ColBERT-v2 drops by 29.2\% and ConstBERT by 7.0\% relative to their zero-shot baselines. The stability of this degradation across seeds indicates this is not a training artifact but a fundamental rejection of the task.

To determine whether this was a simple case of data scarcity, we tripled the training data (428 queries) and applied strict regularization. The result remained identical: the performance still decreased by 16--17\%. Most importantly, early stopping was triggered almost immediately (steps 100--200). This suggests that the model's MS-MARCO-trained priors are not just strong but that the structural mismatch of ToT provides no ``gradient path'' for improvement.

\subsubsection{The Architectural Adaptation Ceiling}

The failure of adaptation confirms that the bottleneck is not the quality of the representations, but the MaxSim operator itself. Because MaxSim’s uniform weighting is hardcoded, the model cannot learn to ignore the 70\% of filler words in a ToT narrative.

For the model to succeed, it would have to learn to map filler words (e.g., ``maybe'') to a null-vector space, an indirect and inefficient workaround. Our results suggest that you cannot fix a scoring error with representation learning. The uniform summation treats every token as a first-class citizen, ensuring that the relevance signal remains diluted regardless of how much data is provided.

\subsubsection{Takeaway}

Fine-tuning is a cure for topical ignorance, not structural inadequacy. When queries differ fundamentally in format and ambiguity, additional training data provides no benefit; in fact, it introduces noise that degrades existing zero-shot capabilities. This establishes an architectural adaptation ceiling: MaxSim's inability to weight tokens discriminatively renders it ``unlearnable'' for long narrative structures. Practitioners should prioritize architectural modifications, such as learned token weighting or hybrid scoring, over data collection when facing structural distribution shifts. 

While the TREC ToT dataset is smaller than MS-MARCO, the observed 89\% performance collapse and the immediate activation of early stopping suggest a structural rejection of the query format rather than a data scarcity artifact. Furthermore, our exact MaxSim analysis (Section~\ref{sec:rq4}) confirms that the failure is rooted in the scoring function's mathematical treatment of long-form signal, a property that remains invariant to dataset size.

%% file: tex_files/tables/finetuning.tex
\begin{table*}[t]
\caption{Fine-tuning results on ToT across all evaluation splits. Both 
models show degradation when fine-tuned on the small TRAIN split (143 
queries)—a counterintuitive result suggesting limited training data and 
distributional mismatch prevent effective adaptation. TEST results are 
multi-seed (n=3, seeds 42/123/456); DEV splits are single-run from 
fine-tuned models.}
\label{tab:finetuning}
\begin{tabular}{llccccr}
\toprule
\textbf{Model} & \textbf{Split} & \textbf{Queries} & \textbf{MRR@10} & \textbf{nDCG@10} & \textbf{R@1000} & \textbf{$\Delta$ Base} \\
\midrule
\multirow{5}{*}{ColBERT-v2} & Base (pretrained) & --- & 5.66\% & 6.07\% & 28.94\% & --- \\
& DEV1 (fine-tuned) & 142 & 3.65\% & 3.78\% & 12.68\% & --35.5\% \\
& DEV2 (fine-tuned) & 143 & 2.23\% & 2.33\% & 13.29\% & --- \\
& DEV3 (fine-tuned) & 536 & 2.43\% & 2.71\% & 10.82\% & --- \\
& TEST (n=3) & 622 & \textbf{4.01\% $\pm$ 0.08\%} & --- & \textbf{21.70\% $\pm$ 0.23\%} & \textbf{--29.2\%} \\
\midrule
\multirow{5}{*}{ConstBERT} & Base (pretrained) & --- & 4.27\% & 4.82\% & 25.72\% & --- \\
& DEV1 (fine-tuned) & 142 & 1.58\% & 1.71\% & 16.90\% & --62.9\% \\
& DEV2 (fine-tuned) & 143 & 0.57\% & 0.64\% & 18.88\% & --86.6\% \\
& DEV3 (fine-tuned) & 536 & 2.37\% & 2.68\% & 17.72\% & --44.5\% \\
& TEST (n=3) & 622 & \textbf{3.97\% $\pm$ 0.15\%} & \textbf{4.48\% $\pm$ 0.14\%} & \textbf{25.35\% $\pm$ 0.77\%} & \textbf{--7.0\%} \\
\midrule
\multicolumn{2}{l}{\textit{MS-MARCO (reference)}} & 6,980 & \textit{39.15\%} & \textit{---} & \textit{90.30\%} & --- \\
\bottomrule
\end{tabular}
\end{table*}

%% file: tex_files/sections/conclusion.tex
\section{Conclusion}
\label{sec:conclusion}

We assessed multi-vector retrieval reproducibility across five dimensions: implementation correctness, backend robustness, domain generalization, structural generalization, and adaptation potential. Our central finding: models can reproduce numerically while failing architecturally—a critical distinction for both reproducibility methodology and deployment practice.

\textbf{Key Findings.} ConstBERT reproduces in-domain effectiveness within 0.05\% MRR@10 (38.99\% vs 39.04\%), confirming implementation correctness. However, undocumented PLAID parameters create an 8-point reproduction gap (31.09\% vs 39.04\%). Centroid coverage analysis reveals why: ConstBERT's 32 vectors map to only 12.1 unique centroids on average, causing sparse footprints that break standard indexing configurations.

On BEIR, ConstBERT degrades systematically (--9.4\% to --17.5\%) while ColBERT-v2 reproduces successfully ($\pm$1.4\%), demonstrating architecture-dependent brittleness. On TREC ToT 2025's long queries (121 vs 6 words), both models collapse 86--97\% (4--5\% MRR@10). Three ablations isolate the cause: (1) query length—performance plateaus beyond 20 words; (2) exact MaxSim—achieves only 5.08\% even without approximation; (3) fine-tuning—degrades performance (--16\% to --29\%) even with 3$\times$ more data. MaxSim's uniform token weighting cannot handle queries where 60--70\% of tokens are filler words.

\textbf{Contributions.} We establish that reproducibility must validate architectural robustness, not just numerical accuracy. Our framework reveals \emph{what} is reproduced (benchmark-optimized vs generalizable) matters as much as \emph{whether}. We identify query structure as an independent failure mode distinct from domain shift, exposing a critical gap in current benchmarks. We provide mechanistic explanations—centroid sparsity, length saturation, architectural ceilings—proving where and why systems fail.

\textbf{Implications.} Backend configurations are first-class reproducibility requirements—undocumented parameters prevent reproduction even when implementations are correct. Benchmark success (MS-MARCO: 39\%, BEIR: 47--50\%) provides insufficient evidence for deployment on structurally different queries. Architectural mismatches cannot be fixed by fine-tuning—when scoring functions break, more data cannot help.

Our findings challenge a fundamental assumption in neural IR: that models achieving state-of-the-art performance on standard benchmarks are production-ready. The field has optimized for short, well-formed queries while real-world information needs span diverse lengths and structures. Until we design retrieval architectures that adaptively weight query tokens and validate them across the full spectrum of query distributions, we risk deploying systems that excel in evaluation yet fail in practice.

%% file: references.bib
@inproceedings{khattab2020colbert,
author = {Khattab, Omar and Zaharia, Matei},
title = {ColBERT: Efficient and Effective Passage Search via Contextualized Late Interaction over BERT},
year = {2020},
isbn = {9781450380164},
publisher = {Association for Computing Machinery},
address = {New York, NY, USA},
url = {https://doi.org/10.1145/3397271.3401075},
doi = {10.1145/3397271.3401075},
booktitle = {Proceedings of the 43rd International ACM SIGIR Conference on Research and Development in Information Retrieval},
pages = {39–48},
numpages = {10},
location = {Virtual Event, China},
series = {SIGIR '20}
}

@article{SA202140,
title = {Examining User Perception and Usage of Voice Search},
journal = {Data and Information Management},
volume = {5},
number = {1},
pages = {40-47},
year = {2021},
issn = {2543-9251},
doi = {10.2478/dim-2020-0046},
url = {https://www.sciencedirect.com/science/article/pii/S2543925122000195},
author = {Ning Sa and Xiaojun (Jenny) Yuan}
}

@article{izacard2021towards,
  title={Towards Unsupervised Dense Information Retrieval with Contrastive Learning},
  author={Izacard, Gautier and Caron, Mathilde and Hosseini, Lucas and Riedel, Sebastian and Bojanowski, Piotr and Joulin, Armand and Grave, Edouard},
  journal={arXiv preprint arXiv:2112.09118},
  year={2021}
}

@inproceedings{gupta2015information,
author = {Gupta, Manish and Bendersky, Michael},
title = {Information Retrieval with Verbose Queries},
year = {2015},
isbn = {9781450336215},
publisher = {Association for Computing Machinery},
address = {New York, NY, USA},
url = {https://doi.org/10.1145/2766462.2767877},
doi = {10.1145/2766462.2767877},
booktitle = {Proceedings of the 38th International ACM SIGIR Conference on Research and Development in Information Retrieval},
pages = {1121–1124},
numpages = {4},
location = {Santiago, Chile},
series = {SIGIR '15}
}

@inproceedings{qian-dou-2022-explicit,
    title = "Explicit Query Rewriting for Conversational Dense Retrieval",
    author = "Qian, Hongjin  and
      Dou, Zhicheng",
    editor = "Goldberg, Yoav  and
      Kozareva, Zornitsa  and
      Zhang, Yue",
    booktitle = "Proceedings of the 2022 Conference on Empirical Methods in Natural Language Processing",
    month = dec,
    year = "2022",
    address = "Abu Dhabi, United Arab Emirates",
    publisher = "Association for Computational Linguistics",
    url = "https://aclanthology.org/2022.emnlp-main.311/",
    doi = "10.18653/v1/2022.emnlp-main.311",
    pages = "4725--4737"
}

@inproceedings{poddar2025learning,
author = {Podder, Dipannita and Paik, Jiaul and Mitra, Pabitra},
title = {Learning Query Token Importance for Effective Document Retrieval with Verbose Queries},
year = {2025},
isbn = {9798400713187},
publisher = {Association for Computing Machinery},
address = {New York, NY, USA},
url = {https://doi.org/10.1145/3734947.3734954},
doi = {10.1145/3734947.3734954},
booktitle = {Proceedings of the 16th Annual Meeting of the Forum for Information Retrieval Evaluation},
pages = {67–75},
numpages = {9},
series = {FIRE '24}
}

@inproceedings{yu2021improving,
author = {Yu, HongChien and Xiong, Chenyan and Callan, Jamie},
title = {Improving Query Representations for Dense Retrieval with Pseudo Relevance Feedback},
year = {2021},
isbn = {9781450384469},
publisher = {Association for Computing Machinery},
address = {New York, NY, USA},
url = {https://doi.org/10.1145/3459637.3482124},
doi = {10.1145/3459637.3482124},
booktitle = {Proceedings of the 30th ACM International Conference on Information \& Knowledge Management},
pages = {3592–3596},
numpages = {5},
location = {Virtual Event, Queensland, Australia},
series = {CIKM '21}
}

@inproceedings{santhanam2022colbertv2,
  title={Colbertv2: Effective and efficient retrieval via lightweight late interaction},
  author={Santhanam, Keshav and Khattab, Omar and Saad-Falcon, Jon and Potts, Christopher and Zaharia, Matei},
  booktitle={Proceedings of the 2022 Conference of the North American Chapter of the Association for Computational Linguistics: Human Language Technologies},
  pages={3715--3734},
  year={2022}
}

@inproceedings{macavaney2025constbert,
  title={Efficient Constant-Space Multi-vector Retrieval},
  author={MacAvaney, Sean and Mallia, Antonio and Tonellotto, Nicola},
  booktitle={European Conference on Information Retrieval},
  pages={237--245},
  year={2025},
  organization={Springer}
}

@inproceedings{santhanam2022plaid,
  title={PLAID: an efficient engine for late interaction retrieval},
  author={Santhanam, Keshav and Khattab, Omar and Potts, Christopher and Zaharia, Matei},
  booktitle={Proceedings of the 31st ACM International Conference on Information \& Knowledge Management},
  pages={1747--1756},
  year={2022}
}

@inproceedings{donabauer2025legal,
  title={A Reproducibility Study of Graph-Based Legal Case Retrieval},
  author={Donabauer, Gregor and Kruschwitz, Udo},
  booktitle={Proceedings of the 48th International ACM SIGIR Conference on Research and Development in Information Retrieval},
  pages={3135--3144},
  year={2025}
}

@inproceedings{yao2025pretraining,
  title={Pre-training vs. Fine-tuning: A Reproducibility Study on Dense Retrieval Knowledge Acquisition},
  author={Yao, Zheng and Wang, Shuai and Zuccon, Guido},
  booktitle={Proceedings of the 48th International ACM SIGIR Conference on Research and Development in Information Retrieval},
  pages={3276--3285},
  year={2025}
}

@inproceedings{ferrari2019worrying,
  author    = {Maurizio Ferrari Dacrema and Paolo Cremonesi and Dietmar Jannach},
  title     = {Are We Really Making Much Progress? {A} Worrying Analysis of Recent Neural Recommendation Approaches},
  booktitle = {Proceedings of the 13th ACM Conference on Recommender Systems (RecSys '19)},
  year      = {2019},
  publisher = {ACM},
  pages     = {101--109},
  doi       = {10.1145/3298689.3347058}
}

@article{thakur2021beir,
  title={Beir: A heterogenous benchmark for zero-shot evaluation of information retrieval models},
  author={Thakur, Nandan and Reimers, Nils and R{\"u}ckl{\'e}, Andreas and Srivastava, Abhishek and Gurevych, Iryna},
  journal={arXiv preprint arXiv:2104.08663},
  year={2021}
}

@article{trec2025tot,
  title={Overview of the TREC 2025 Tip-of-the-Tongue track},
  author={Arguello, Jaime and Diaz, Fernando and Fr{\"o}ebe, Maik and Kim, To Eun and Mitra, Bhaskar},
  journal={arXiv preprint arXiv:2601.20671},
  year={2026}
}

@article{johnson2019billion,
  title={Billion-scale similarity search with GPUs},
  author={Johnson, Jeff and Douze, Matthijs and J{\'e}gou, Herv{\'e}},
  journal={IEEE Transactions on Big Data},
  volume={7},
  number={3},
  pages={535--547},
  year={2019},
  publisher={IEEE}
}
